\documentclass[12pt]{article}

\usepackage[utf8]{inputenc}
\usepackage[T1]{fontenc}
\usepackage[standard]{ntheorem}

\usepackage[english]{babel}
\usepackage{amsmath}
\usepackage{amssymb}
\usepackage{graphicx}
\usepackage{hyperref}

\usepackage{url}

\def\shift{\textit{shift}}
\def\shi{\textit{sh}}
\def\sh1{\textit{sh1}}

\def\pp{\mathinner{\ldotp\ldotp}}	

\begin{document}

\title{Optimal-Hash Exact String Matching Algorithms}

\author{Thierry Lecroq\footnote{CURIB,
UFR des Sciences et des Techniques,
Université de Rouen Normandie,
76821 Mont-Saint-Aignan Cedex,
France,
\protect\url{Thierry.Lecroq@univ-rouen.fr},
\protect\url{monge.univ-mlv.fr/~lecroq/lec_en.html}
}
}

\date{Univ Rouen Normandie, INSA Rouen Normandie, Université Le Havre Normandie, Normandie Univ, LITIS UR 4108, CNRS NormaSTIC FR 3638, IRIB, F-76000 Rouen, France}

\maketitle

\begin{abstract}
String matching is the problem of finding all the occurrences of a
 pattern in a text.
We propose improved versions of the fast  family of string matching algorithms based on
 hashing $q$-grams.
The improvement consists of considering minimal values $q$ such that each
 $q$-grams of the pattern has a unique hash value.
The new algorithms are fastest than algorithm of the HASH family for short patterns
 on large size alphabets.
\end{abstract}

\section{Introduction}

The string matching problem consists of finding one or more usually
 all the occurrences of a pattern $x=x[0\pp m-1]$ of length $m$ in a text $y=y[0\pp n-1]$
 of length $n$.
It can occur in many applications, for instance in information retrieval, bibliographic search and
 molecular biology.
It has been extensively studied and numerous techniques and algorithms
 have been designed to solve this problem
 (see~\cite{CL2004,Faro2011a}).
We are interested here in the problem where the pattern is given first
 and can then be searched in various texts.
Thus a preprocessing phase is allowed on the pattern.

Basically a string-matching algorithm uses a window to scan the text.
The size of this window is equal to the length of the pattern.
It first aligns the left ends of the window and the text.
Then it checks if the pattern occurs in the window (this specific work
 is called an {\em attempt}) and {\em shifts} the window to the right.
It repeats the same procedure again until the right end of the window
 goes beyond the right end of the text.
The many different solutions differ in the way they compare the content
 of the window and the pattern, in the way they compute the lengths of the shifts and
 in the information that is stored from one attempt to the next.

There is no universal algorithm in the sense that the efficiency of a particular algorithm
 mainly depends on the size of the alphabet and the length of the pattern~\cite{FL2010}.
According to~\cite{Faro2011a}, the many algorithms can be classified in character comparison
 based algorithms,
automata based algorithms and bit-parallelism based algorithms.

Among character comparison based algorithms, the HASH family of string matching algorithms~\cite{Lec2007}
 consists basically of hashing every $q$-grams $u$ of $x$ resulting in $h(u)$ 
 and computing a shift function for
 $h(u)$ equal to $\min\{m-i \mid h(x[i\pp i+q-1]=h(u)\}$.
The other feature of the algorithms is that they avoid loops for computing $h(u)$ thus
 $q$ is not a parameter of the algorithm, there should be one algorithm per value of $q$. 
 
Since its publication in 2007 the HASH family of string matching algorithms
 has aroused a lot of interest and has still been considered among the fastest
 algorithms~\cite{Nguyen2019, Karcio2021, Alssul2017, Pelto2014, Alssul2022, Tahir2017}
  and even the fastest ones in some special cases~\cite{aa2015,Zavad2017,Koba2020}.
However, it has a main drawback: the length $q$ of the $q$-grams has to be determined in advance
 which implies that the algorithm does not work for pattern of length smaller than $q$.
We propose, in this article, algorithms that remedy to this problem
 by selecting the length of the $q$-gram as the smallest $q$
 such that the pattern has no two identical substrings of length $q$ having the same hash value.
It has three advantages: first the algorithms can considered pattern of any length,
 second when $q=1$ or $q=2$ it is easily possible to use perfect hashing
 and avoid some character comparisons when checking candidates
 and third they are faster in some cases than the original HASH algorithms.
 
This article is organized as follows: Section~\ref{sect-algo} presents
 the new algorithms, Section~\ref{sect-expe} shows
 experimental results and Section~\ref{sect-conc} provides our
 conclusion.

\section{\label{sect-algo}The new algorithms}

The HASH family string matching algorithms consider substrings of length $q$.
Substrings $u$ of the pattern $x$ of such a length are hashed using a function $h$ into integer values
 within 0 and $S$.
Then a shift value is defined for every hash value as follows.
For $0\le c \le S$:
$$\shift[c]=\begin{cases}
m-1-i & \mbox{with } i=\max\{0\le j \le m-q+1 \mid h(x[j\pp j+q-1]) = c\}\cr
m-q & \mbox{when such an $i$ does not exist}\cr
\end{cases}$$
where $S$ is the size of the hash table.

The searching phase of the algorithm consists in reading substrings $u$ of length $q$.
If $\shift[h(u)] > 0$ then a shift of length $\shift[h(u)]$ is applied.
Otherwise, when $\shift[h(u)] = 0$ the pattern $x$ is naively checked in the text.
In this case a shift of length $\shi$ is applied where
 $\shi=m-1-i$ with $i=\max\{0\le j \le m-q \mid h(x[j\pp j+q-1]) = h(x[m-q+1\pp m-1]\}$.

The idea of the new algorithms is to consider substrings of length $q$
 when the pattern has no two substrings of length $q$ hashed to same the value.
This can be easily computed using the suffix array of the pattern along
 with the Longest Common Prefix array~\cite{CHL2007,CLR2021}.
This gives a lower bound of $q$ and then all substrings of $x$ have to be
 hashed to check that all the hash values are different.
If not $q$ is incremented by one and the values are checked again until all
 the hash values of the $q$-gram of $x$ are different.
 
When $q=1$ then it is possible to use easily perfect hashing with a
 hashing table of size $256$ and when $q=2$ the size of the hash table
 grows to $65536$.
Tables of this size could efficiently be allocated on the execution stack.
For larger values of $q$, simple perfect hashing requires larger hash
 tables that cannot be allocated on the execution stack and that needs to
 be allocated and deallocated explicitly in the heap which considerably slows down
 the algorithm.

The advantage of perfect hashing is that if $h(y[j\pp j+q-1])=0$ (for some $0\le j \le n-q$)
 then it means that
 $y[j\pp j+q-1]=x[m-q\pp m-1]$ thus the checking phase only requires to test
 if $y[j+q-m\pp j-1] = x[0\pp m-q-1]$ whereas if the hashing is not perfect
 the test becomes $y[j+q-m\pp j+q-1] = x$.

\section{\label{sect-expe}Experimental results}

To evaluate the efficiency of the new string matching algorithms
 we used the SMART system~\cite{Faro2016}.
We perform several
 experiments with different algorithms on different data sets of the system.

\subsection{Algorithms}

We have tested 6 algorithms:
\begin{itemize}
\item
HASH3~\cite{Lec2007};
\item
HASH5~\cite{Lec2007};
\item
HASH8~\cite{Lec2007};
\item 
OHASH1: perfect hashing for $q=1$, hashing with possible collisions for $2\le q \le 10$ and HASH8 when $q>10$; 
\item
OHASH2: perfect hashing for $q=1$ and $q=2$, hashing with possible collisions for $3\le q \le 10$ and HASH8 when $q>10$;
\item
OHASH3: perfect hashing for $q=1$ and $q=2$, HASH3 for $3\le q \le 10$ and HASH8 when $q>10$.
\end{itemize}

These algorithms have been coded in C in an homogeneous way to keep the
 comparison significant.
Since the suffix array has to be build for the pattern which is relatively short comparing to the text,
 its construction has been naively implemented.
The machine we used has an Intel Xeon processor at 2.4GHz
 running Ubuntu version 14.04.5 LTS.
The code of the OHASH algorithms is available on \url{github.com/lecroq/ohash}.

\subsection{Data}

We give experimental results for the running times of the above algorithms for
 different types of text: random texts alphabet of size 8, 16, 32, 64, 128 and 250,
 texts in natural languages (English and Italian) and protein sequences.
We consider short patterns (even length within 2 and 22).

\subsection{Results}

\begin{figure}
\begin{center}
\includegraphics[angle=-90,width=12cm]{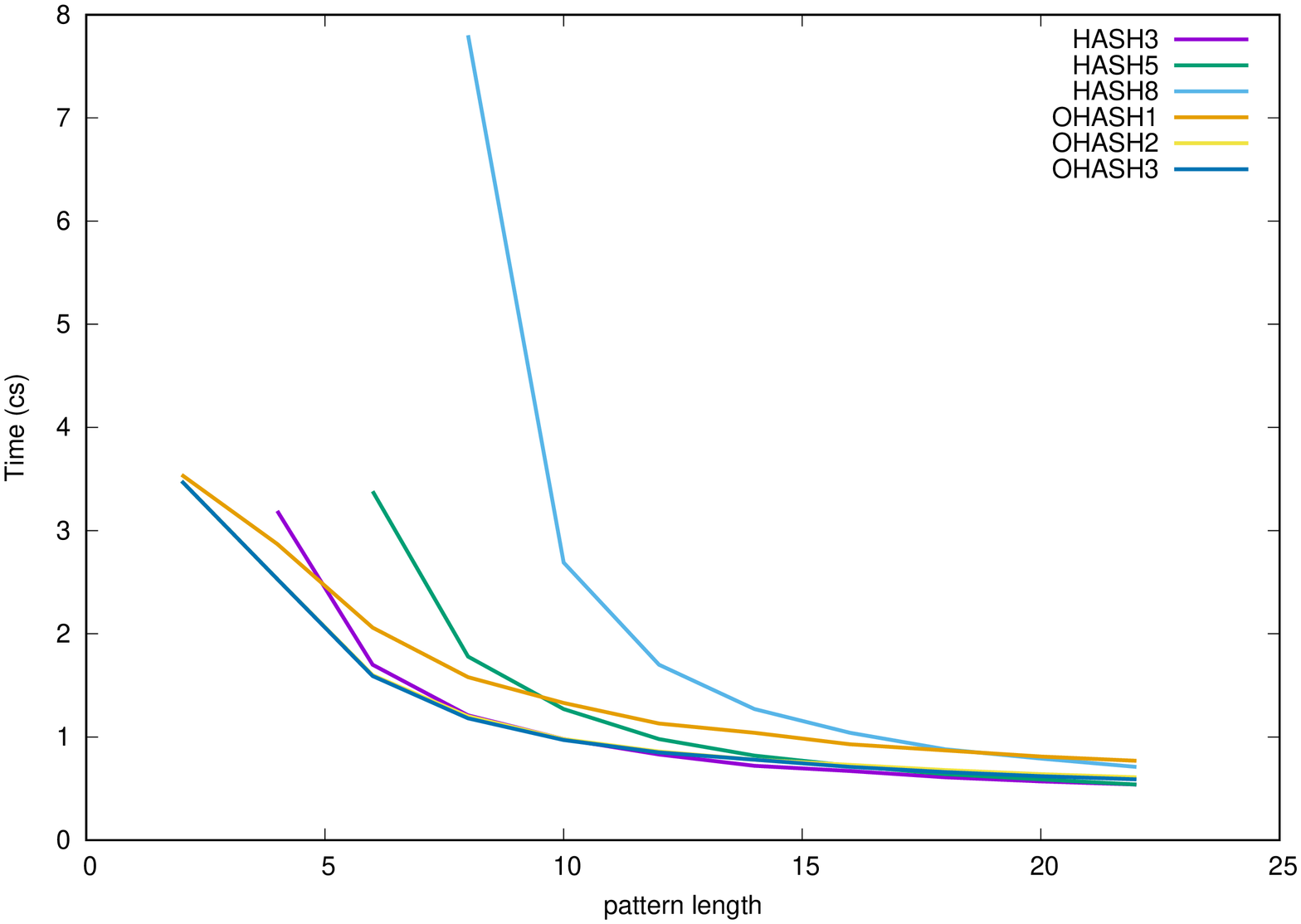}
\end{center}
\caption{\label{figu-rand8}Results for short patterns and random text on an alphabet of size 8}
\end{figure}

\begin{figure}
\begin{center}
\includegraphics[angle=-90,width=12cm]{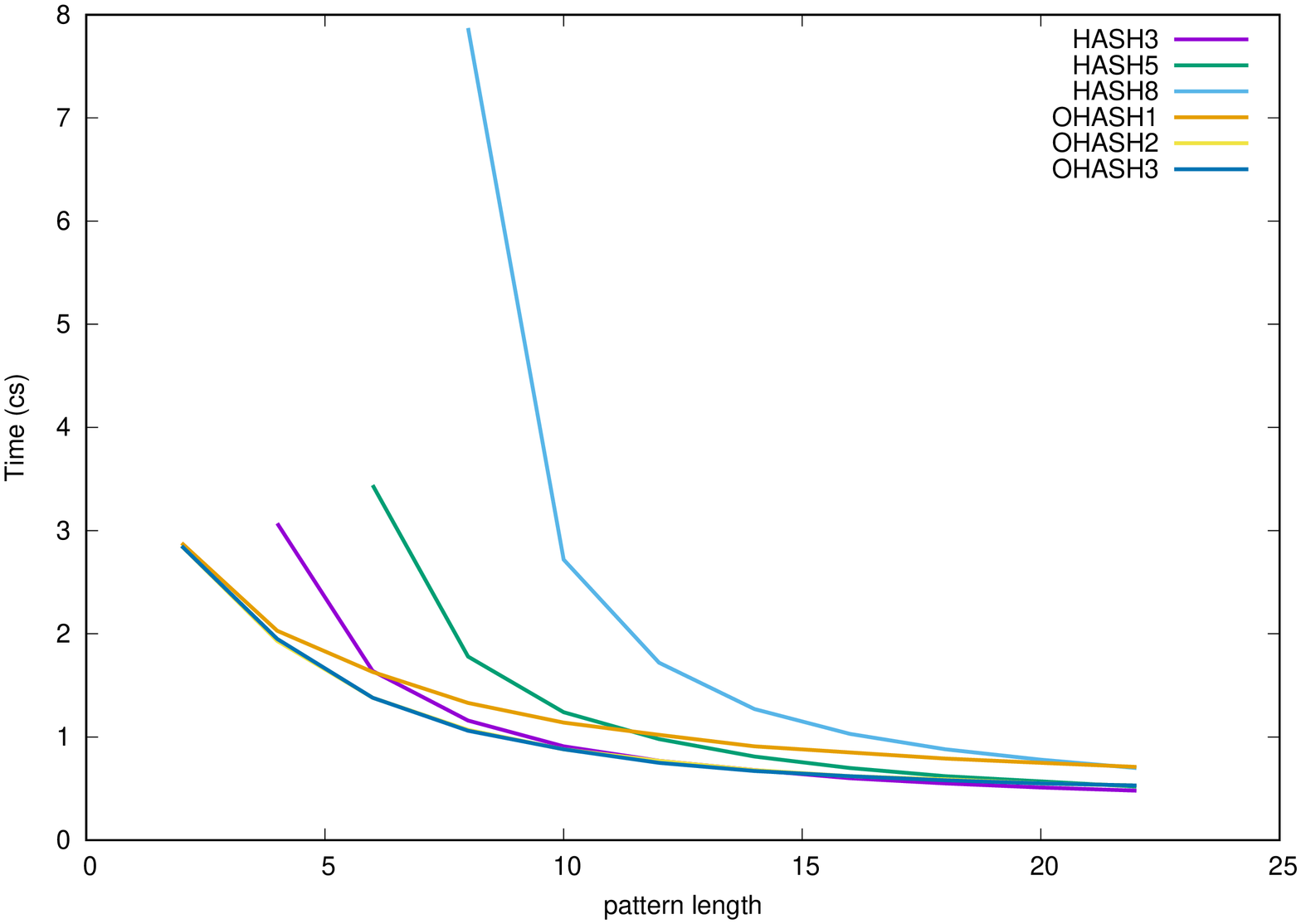}
\end{center}
\caption{\label{figu-rand16}Results for short patterns and random text on an alphabet of size 16}
\end{figure}

\begin{figure}
\begin{center}
\includegraphics[angle=-90,width=12cm]{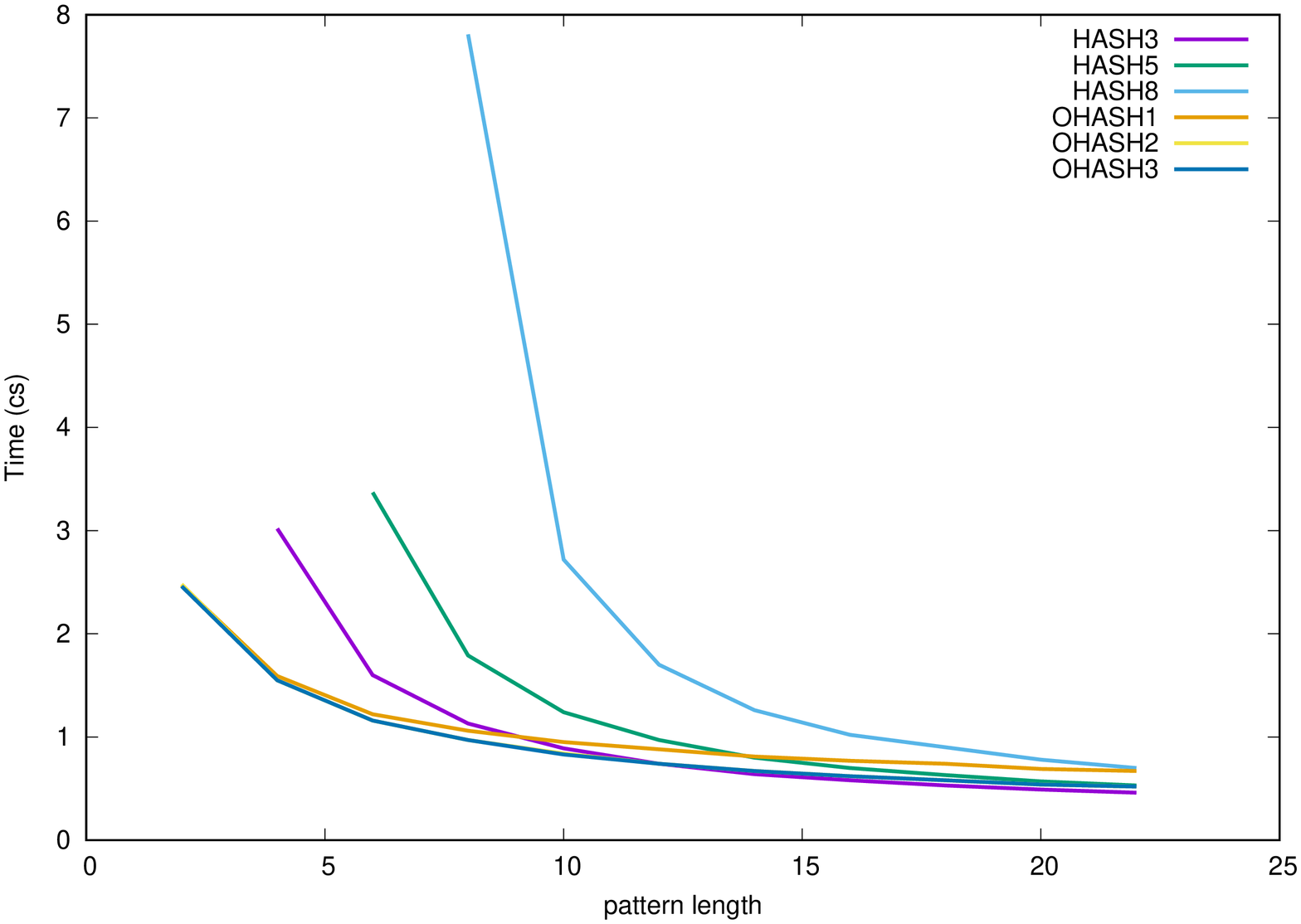}
\end{center}
\caption{\label{figu-rand32}Results for short patterns and random text on an alphabet of size 32}
\end{figure}

\begin{figure}
\begin{center}
\includegraphics[angle=-90,width=12cm]{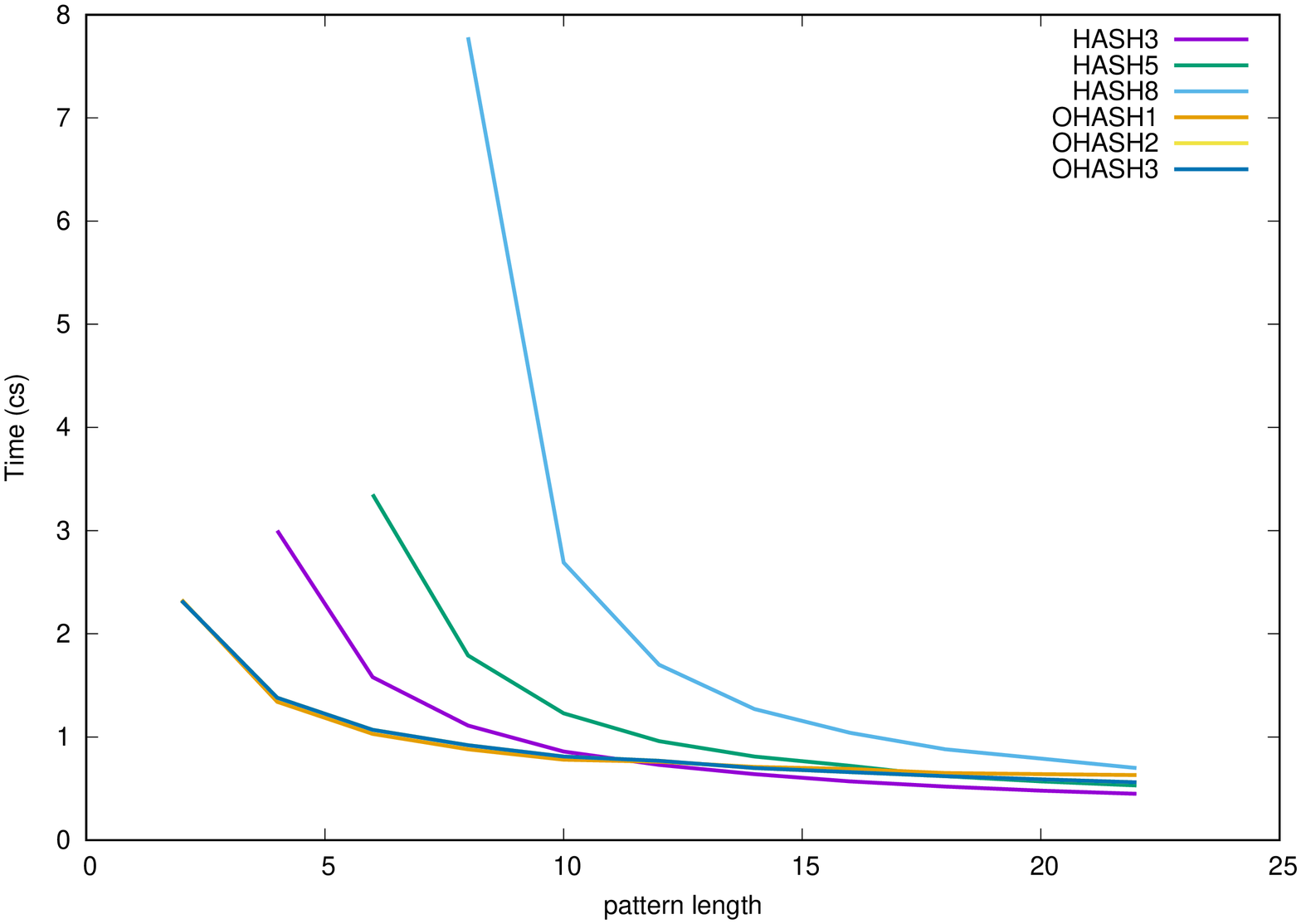}
\end{center}
\caption{\label{figu-rand64}Results for short patterns and random text on an alphabet of size 64}
\end{figure}

\begin{figure}
\begin{center}
\includegraphics[angle=-90,width=12cm]{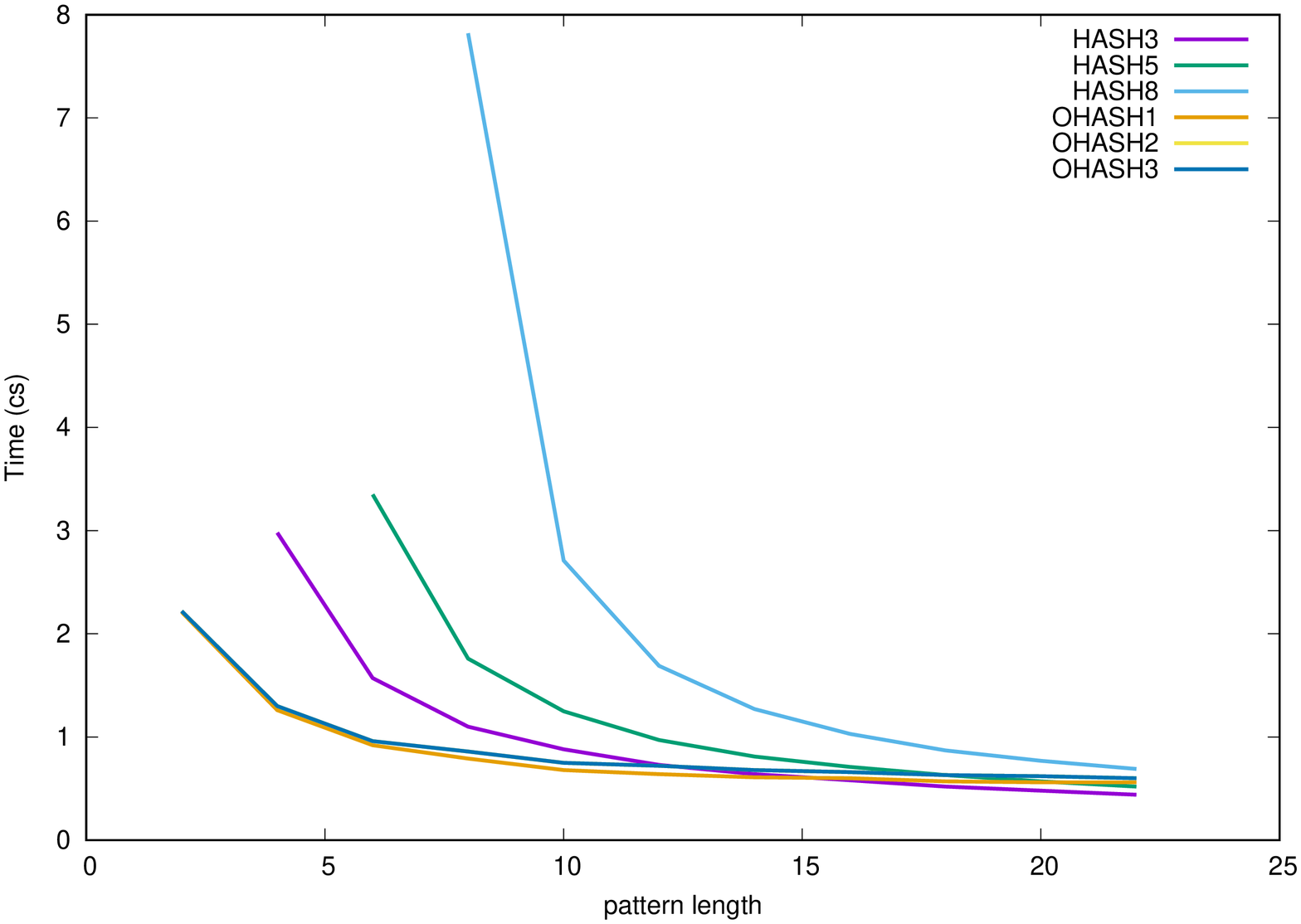}
\end{center}
\caption{\label{figu-rand128}Results for short patterns and random text on an alphabet of size 128}
\end{figure}

\begin{figure}
\begin{center}
\includegraphics[angle=-90,width=12cm]{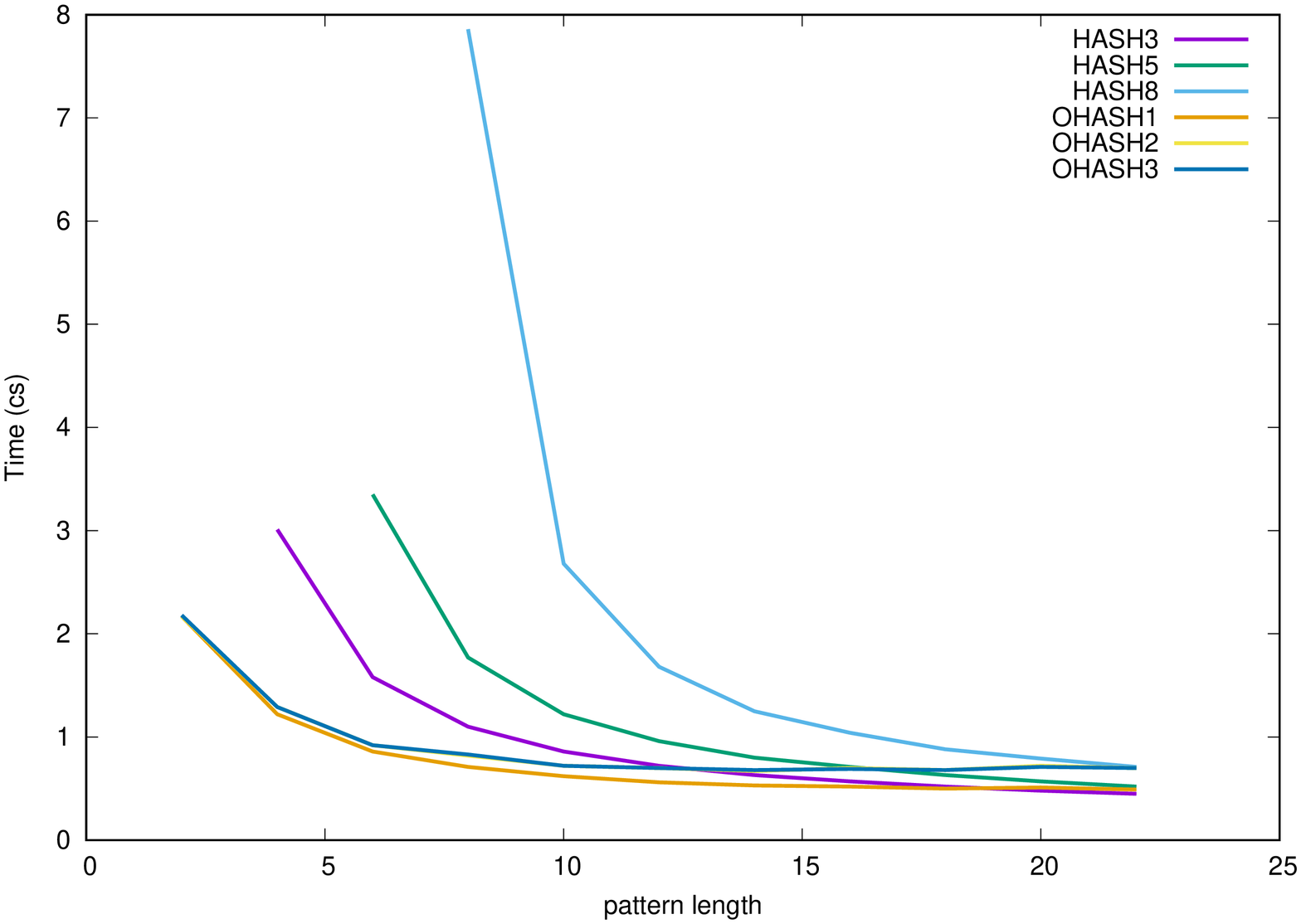}
\end{center}
\caption{\label{figu-rand250}Results for short patterns and random text on an alphabet of size 250}
\end{figure}

\begin{figure}
\begin{center}
\includegraphics[angle=-90,width=12cm]{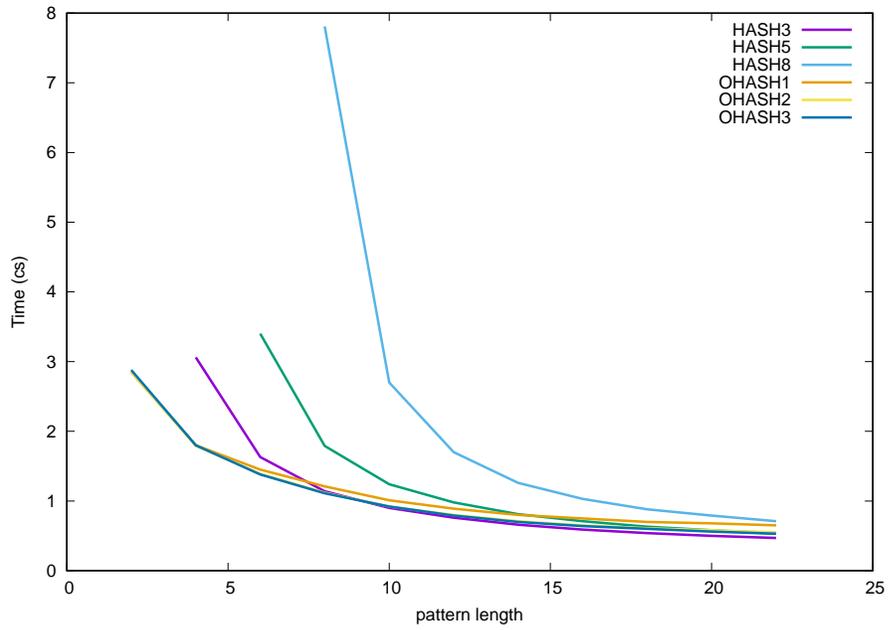}
\end{center}
\caption{\label{figu-eng}Results for short patterns on English texts}
\end{figure}

\begin{figure}
\begin{center}
\includegraphics[angle=-90,width=12cm]{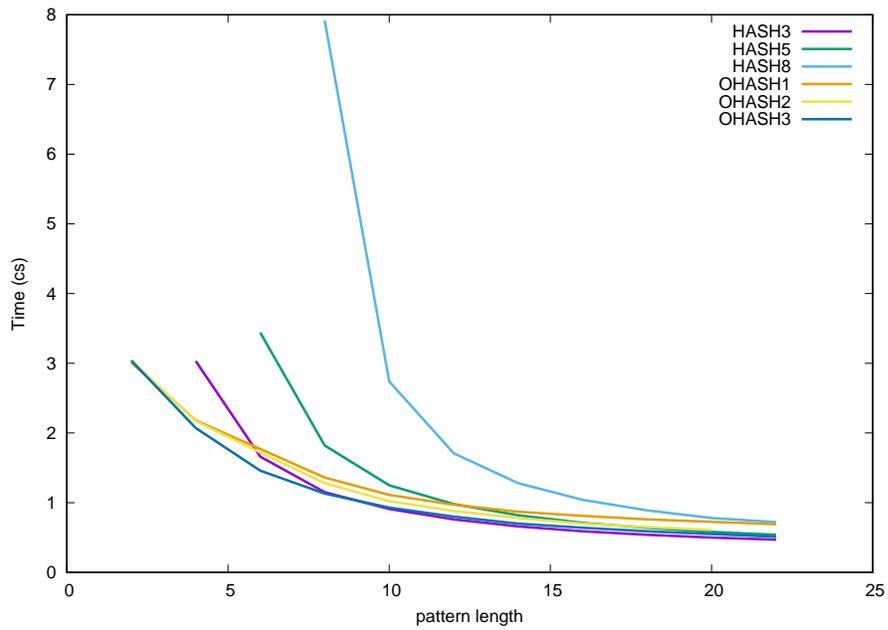}
\end{center}
\caption{\label{figu-ita}Results for short patterns on Italian texts}
\end{figure}

\begin{figure}
\begin{center}
\includegraphics[angle=-90,width=12cm]{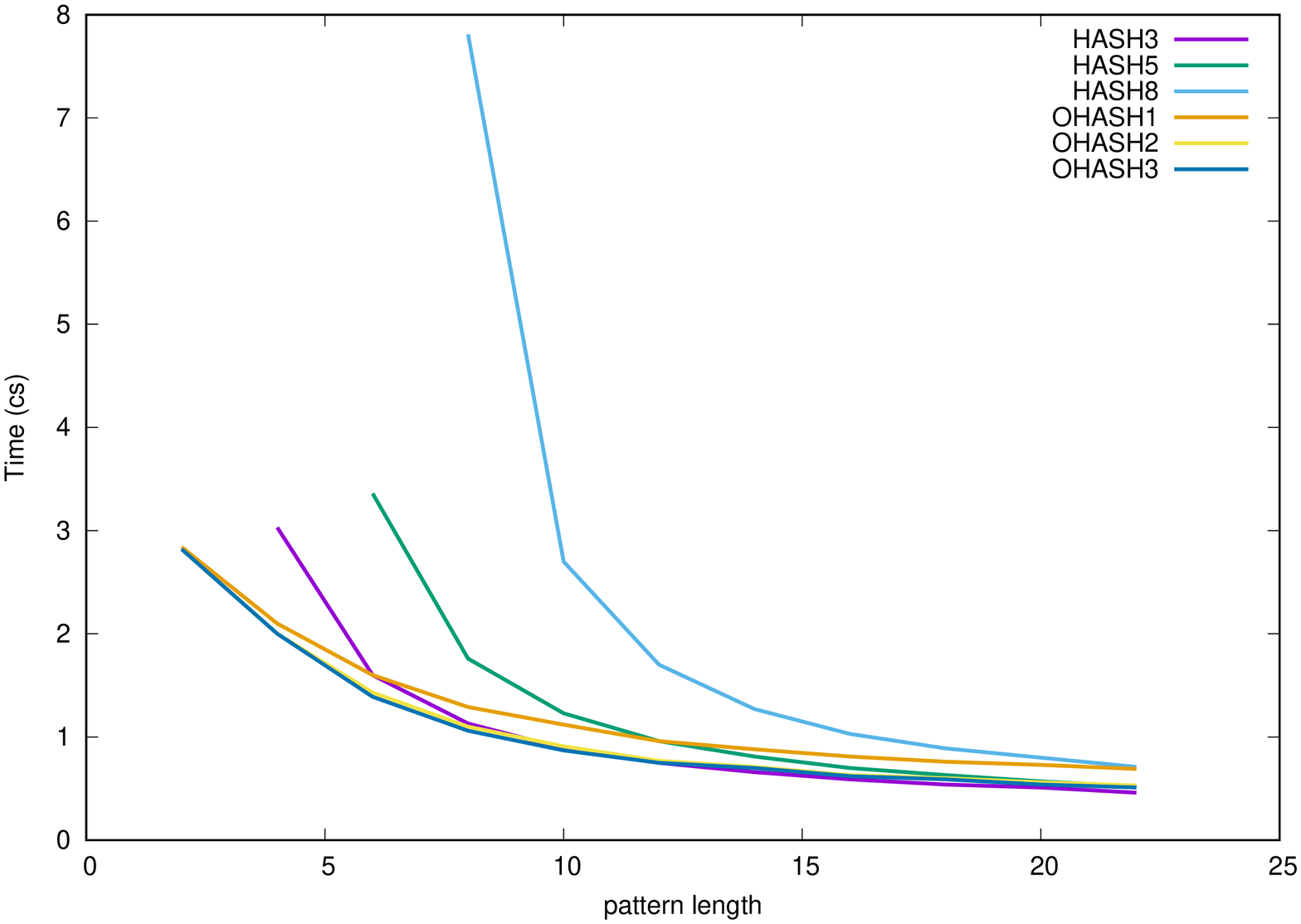}
\end{center}
\caption{\label{figu-prot}Results for short patterns on protein sequences}
\end{figure}

The results for short patterns (length less or equal to 22)
 are presented in tables \ref{resu-rand8} to \ref{resu-prot}
 (fastest results are in bold face)
 and in figures \ref{figu-rand8} to \ref{figu-prot}.

\begin{table}
\caption{\label{resu-rand8}Results for short patterns and random text on an alphabet of size 8}
\begin{center}
\setlength{\tabcolsep}{3pt}
\begin{tabular}{|l|c|c|c|c|c|c|c|c|c|c|c|}
\hline
$m$ & $2$ & $4$ & $6$ & $8$ & $10$ & $12$ & $14$ & $16$ & $18$ & $20$ & $22$\\
\hline
\textsc{HASH3} & - & 3.19 & 1.70 & 1.21 & 0.98 & \textbf{0.83} & \textbf{0.72} & \textbf{0.67} & \textbf{0.61} & \textbf{0.57} & \textbf{0.54}\\
\textsc{HASH5} & - & - & 3.38 & 1.78 & 1.27 & 0.98 & 0.82 & 0.72 & 0.64 & 0.59 & \textbf{0.54}\\
\textsc{HASH8} & - & - & - & 7.80 & 2.69 & 1.70 & 1.27 & 1.04 & 0.88 & 0.79 & 0.71\\
\textsc{OHASH1} & 3.54 & 2.87 & 2.06 & 1.58 & 1.33 & 1.13 & 1.04 & 0.93 & 0.87 & 0.81 & 0.77\\
\textsc{OHASH2} & \textbf{3.48} & \textbf{2.53} & 1.60 & 1.20 & 0.98 & 0.86 & 0.78 & 0.73 & 0.68 & 0.64 & 0.61\\
\textsc{OHASH3} & \textbf{3.48} & \textbf{2.53} & \textbf{1.59} & \textbf{1.18} & \textbf{0.97} & 0.85 & 0.78 & 0.71 & 0.66 & 0.62 & 0.59\\
\hline
\end{tabular}

\end{center}
\end{table}

\begin{table}
\caption{\label{resu-rand16}Results for short patterns and random text on an alphabet of size 16}
\begin{center}
\setlength{\tabcolsep}{3pt}
\begin{tabular}{|l|lllllllllll|}
\hline
$m$ & $2$ & $4$ & $6$ & $8$ & $10$ & $12$ & $14$ & $16$ & $18$ & $20$ & $22$\\
\hline
\textsc{HASH3} & - & 3.07 & 1.64 & 1.16 & 0.91 & 0.77 & 0.68 & \textbf{0.60} & \textbf{0.55} & \textbf{0.51} & \textbf{0.48}\\
\textsc{HASH5} & - & - & 3.44 & 1.78 & 1.24 & 0.98 & 0.81 & 0.70 & 0.62 & 0.57 & 0.52\\
\textsc{HASH8} & - & - & - & 7.87 & 2.72 & 1.72 & 1.27 & 1.03 & 0.88 & 0.78 & 0.70\\
\textsc{OHASH1} & 2.88 & 2.03 & 1.63 & 1.33 & 1.14 & 1.02 & 0.91 & 0.85 & 0.79 & 0.75 & 0.71\\
\textsc{OHASH2} & \textbf{2.85} & \textbf{1.93} & \textbf{1.38} & 1.07 & \textbf{0.88} & 0.77 & 0.68 & 0.62 & 0.59 & 0.55 & 0.53\\
\textsc{OHASH3} & \textbf{2.85} & 1.95 & \textbf{1.38} & \textbf{1.06} & \textbf{0.88} & \textbf{0.75} & \textbf{0.67} & 0.62 & 0.58 & 0.55 & 0.53\\
\hline
\end{tabular}

\end{center}
\end{table}

\begin{table}
\caption{\label{resu-rand32}Results for short patterns and random text on an alphabet of size 32}
\begin{center}
\setlength{\tabcolsep}{3pt}
\begin{tabular}{|l|lllllllllll|}
\hline
$m$ & $2$ & $4$ & $6$ & $8$ & $10$ & $12$ & $14$ & $16$ & $18$ & $20$ & $22$\\
\hline
\textsc{HASH3} & - & 3.02 & 1.60 & 1.13 & 0.89 & \textbf{0.74} & \textbf{0.64} & \textbf{0.58} & \textbf{0.53} & \textbf{0.49} & \textbf{0.46}\\
\textsc{HASH5} & - & - & 3.37 & 1.79 & 1.24 & 0.97 & 0.80 & 0.70 & 0.63 & 0.57 & 0.53\\
\textsc{HASH8} & - & - & - & 7.81 & 2.72 & 1.70 & 1.26 & 1.02 & 0.90 & 0.78 & 0.70\\
\textsc{OHASH1} & 2.47 & 1.59 & 1.22 & 1.06 & 0.95 & 0.88 & 0.81 & 0.77 & 0.74 & 0.69 & 0.67\\
\textsc{OHASH2} & 2.48 & \textbf{1.55} & \textbf{1.16} & \textbf{0.97} & 0.84 & \textbf{0.74} & 0.67 & 0.62 & 0.58 & 0.54 & 0.52\\
\textsc{OHASH3} & \textbf{2.46} & \textbf{1.55} & \textbf{1.16} & \textbf{0.97} & \textbf{0.83} & \textbf{0.74} & 0.67 & 0.62 & 0.58 & 0.54 & 0.52\\
\hline
\end{tabular}

\end{center}
\end{table}

\begin{table}
\caption{\label{resu-rand64}Results for short patterns and random text on an alphabet of size 64}
\begin{center}
\setlength{\tabcolsep}{3pt}
\begin{tabular}{|l|lllllllllll|}
\hline
$m$ & $2$ & $4$ & $6$ & $8$ & $10$ & $12$ & $14$ & $16$ & $18$ & $20$ & $22$\\
\hline
\textsc{HASH3} & - & 3.00 & 1.58 & 1.11 & 0.86 & \textbf{0.73} & \textbf{0.64} & \textbf{0.57} & \textbf{0.52} & \textbf{0.48} & \textbf{0.45}\\
\textsc{HASH5} & - & - & 3.35 & 1.79 & 1.23 & 0.96 & 0.81 & 0.72 & 0.62 & 0.57 & 0.53\\
\textsc{HASH8} & - & - & - & 7.78 & 2.69 & 1.70 & 1.27 & 1.04 & 0.88 & 0.79 & 0.70\\
\textsc{OHASH1} & 2.33 & \textbf{1.34} & \textbf{1.03} & \textbf{0.88} & \textbf{0.78} & 0.76 & 0.71 & 0.69 & 0.65 & 0.64 & 0.63\\
\textsc{OHASH2} & \textbf{2.32} & 1.38 & 1.07 & 0.92 & 0.81 & 0.77 & 0.70 & 0.66 & 0.62 & 0.59 & 0.56\\
\textsc{OHASH3} & \textbf{2.32} & 1.38 & 1.07 & 0.92 & 0.81 & 0.77 & 0.70 & 0.66 & 0.62 & 0.59 & 0.56\\
\hline
\end{tabular}

\end{center}
\end{table}

\begin{table}
\caption{\label{resu-rand128}Results for short patterns and random text on an alphabet of size 128}
\begin{center}
\setlength{\tabcolsep}{3pt}
\begin{tabular}{|l|lllllllllll|}
\hline
$m$ & $2$ & $4$ & $6$ & $8$ & $10$ & $12$ & $14$ & $16$ & $18$ & $20$ & $22$\\
\hline
\textsc{HASH3} & - & 2.98 & 1.57 & 1.10 & 0.88 & 0.73 & 0.64 & \textbf{0.58} & \textbf{0.52} & \textbf{0.48} & \textbf{0.44}\\
\textsc{HASH5} & - & - & 3.35 & 1.76 & 1.25 & 0.97 & 0.81 & 0.71 & 0.63 & 0.57 & 0.52\\
\textsc{HASH8} & - & - & - & 7.82 & 2.71 & 1.69 & 1.27 & 1.03 & 0.87 & 0.77 & 0.69\\
\textsc{OHASH1} & \textbf{2.21} & \textbf{1.26} & \textbf{0.92} & \textbf{0.79} & \textbf{0.68} & \textbf{0.64} & \textbf{0.61} & 0.60 & 0.57 & 0.56 & 0.56\\
\textsc{OHASH2} & 2.22 & 1.30 & 0.96 & 0.86 & 0.75 & 0.72 & 0.68 & 0.66 & 0.63 & 0.62 & 0.60\\
\textsc{OHASH3} & 2.22 & 1.30 & 0.96 & 0.86 & 0.75 & 0.72 & 0.68 & 0.66 & 0.63 & 0.62 & 0.60\\
\hline
\end{tabular}

\end{center}
\end{table}

\begin{table}
\caption{\label{resu-rand250}Results for short patterns and random text on an alphabet of size 250}
\begin{center}
\setlength{\tabcolsep}{3pt}
\begin{tabular}{|l|lllllllllll|}
\hline
$m$ & $2$ & $4$ & $6$ & $8$ & $10$ & $12$ & $14$ & $16$ & $18$ & $20$ & $22$\\
\hline
\textsc{HASH3} & - & 3.01 & 1.58 & 1.10 & 0.86 & 0.72 & 0.63 & 0.57 & 0.52 & \textbf{0.48} & \textbf{0.45}\\
\textsc{HASH5} & - & - & 3.35 & 1.77 & 1.22 & 0.96 & 0.80 & 0.71 & 0.63 & 0.57 & 0.52\\
\textsc{HASH8} & - & - & - & 7.86 & 2.68 & 1.68 & 1.25 & 1.04 & 0.88 & 0.79 & 0.71\\
\textsc{OHASH1} & \textbf{2.17} & \textbf{1.22} & \textbf{0.86} & \textbf{0.71} & \textbf{0.62} & \textbf{0.56} & \textbf{0.53} & \textbf{0.52} & \textbf{0.50} & 0.51 & 0.49\\
\textsc{OHASH2} & \textbf{2.17} & 1.29 & 0.92 & 0.82 & 0.72 & 0.70 & 0.67 & 0.70 & 0.68 & 0.72 & 0.70\\
\textsc{OHASH3} & 2.18 & 1.29 & 0.92 & 0.83 & 0.72 & 0.70 & 0.68 & 0.69 & 0.68 & 0.71 & 0.70\\
\hline
\end{tabular}

\end{center}
\end{table}

\begin{table}
\caption{\label{resu-eng}Results for short patterns and English texts}
\begin{center}
\setlength{\tabcolsep}{3pt}
\begin{tabular}{|l|lllllllllll|}
\hline
$m$ & $2$ & $4$ & $6$ & $8$ & $10$ & $12$ & $14$ & $16$ & $18$ & $20$ & $22$\\
\hline
\textsc{HASH3} & - & 3.06 & 1.63 & 1.14 & \textbf{0.90} & \textbf{0.76} & \textbf{0.66} & \textbf{0.59} & \textbf{0.54} & \textbf{0.50} & \textbf{0.47}\\
\textsc{HASH5} & - & - & 3.40 & 1.79 & 1.24 & 0.98 & 0.81 & 0.71 & 0.63 & 0.58 & 0.53\\
\textsc{HASH8} & - & - & - & 7.81 & 2.70 & 1.70 & 1.26 & 1.03 & 0.88 & 0.79 & 0.71\\
\textsc{OHASH1} & \textbf{2.85} & 1.80 & 1.45 & 1.21 & 1.01 & 0.89 & 0.80 & 0.75 & 0.70 & 0.68 & 0.65\\
\textsc{OHASH2} & \textbf{2.85} & \textbf{1.79} & 1.39 & 1.12 & 0.92 & 0.80 & 0.71 & 0.65 & 0.61 & 0.58 & 0.55\\
\textsc{OHASH3} & 2.88 & 1.80 & \textbf{1.38} & \textbf{1.11} & 0.92 & 0.79 & 0.70 & 0.64 & 0.60 & 0.56 & 0.53\\
\hline
\end{tabular}

\end{center}
\end{table}

\begin{table}
\caption{\label{resu-ita}Results for short patterns and Italian Texts}
\begin{center}
\setlength{\tabcolsep}{3pt}
\begin{tabular}{|l|lllllllllll|}
\hline
$m$ & $2$ & $4$ & $6$ & $8$ & $10$ & $12$ & $14$ & $16$ & $18$ & $20$ & $22$\\
\hline
\textsc{HASH3} & - & 3.03 & 1.66 & 1.15 & \textbf{0.91} & \textbf{0.76} & \textbf{0.66} & \textbf{0.59} & \textbf{0.54} & \textbf{0.50} & \textbf{0.47}\\
\textsc{HASH5} & - & - & 3.44 & 1.82 & 1.25 & 0.98 & 0.82 & 0.71 & 0.64 & 0.58 & 0.54\\
\textsc{HASH8} & - & - & - & 7.92 & 2.74 & 1.71 & 1.28 & 1.04 & 0.89 & 0.78 & 0.72\\
\textsc{OHASH1} & \textbf{3.01} & 2.18 & 1.77 & 1.36 & 1.11 & 0.97 & 0.87 & 0.81 & 0.76 & 0.71 & 0.69\\
\textsc{OHASH2} & 3.04 & 2.17 & 1.72 & 1.28 & 1.02 & 0.88 & 0.78 & 0.70 & 0.65 & 0.61 & 0.58\\
\textsc{OHASH3} & 3.04 & \textbf{2.07} & \textbf{1.46} & \textbf{1.13} & 0.93 & 0.80 & 0.70 & 0.64 & 0.59 & 0.55 & 0.51\\
\hline
\end{tabular}

\end{center}
\end{table}

\begin{table}
\caption{\label{resu-prot}Results for short patterns and protein sequences}
\begin{center}
\setlength{\tabcolsep}{3pt}
\begin{tabular}{|l|lllllllllll|}
\hline
$m$ & $2$ & $4$ & $6$ & $8$ & $10$ & $12$ & $14$ & $16$ & $18$ & $20$ & $22$\\
\hline
\textsc{HASH3} & - & 3.03 & 1.60 & 1.13 & 0.89 & \textbf{0.75} & \textbf{0.66} & \textbf{0.59} & \textbf{0.54} & \textbf{0.51} & \textbf{0.46}\\
\textsc{HASH5} & - & - & 3.36 & 1.76 & 1.23 & 0.96 & 0.81 & 0.70 & 0.63 & 0.57 & 0.52\\
\textsc{HASH8} & - & - & - & 7.81 & 2.70 & 1.70 & 1.27 & 1.03 & 0.89 & 0.80 & 0.71\\
\textsc{OHASH1} & 2.84 & 2.10 & 1.60 & 1.29 & 1.12 & 0.96 & 0.88 & 0.81 & 0.76 & 0.73 & 0.69\\
\textsc{OHASH2} & \textbf{2.82} & \textbf{2.00} & 1.43 & 1.10 & 0.91 & 0.77 & 0.71 & 0.63 & 0.60 & 0.56 & 0.53\\
\textsc{OHASH3} & \textbf{2.82} & \textbf{2.00} & \textbf{1.39} & \textbf{1.06} & \textbf{0.87} & \textbf{0.75} & 0.70 & 0.62 & 0.59 & 0.54 & 0.51\\
\hline
\end{tabular}

\end{center}
\end{table}

From the experiments it appears that
algorithm OHASH3 is the fastest:
\begin{itemize}
\item on random texts on an alphabet of size 8 for $2\le m \le 10$;
\item on random texts on an alphabet of size 16 for $2\le m \le 14$;
\item on random texts on an alphabet of size 32 for $2\le m \le 12$;
\item on English texts for $2\le m \le 8$;
\item on Italian texts for $2\le m \le 8$;
\item on protein sequences for $2\le m \le 12$;
\end{itemize}
and that algorithm OHASH1 is the fastest:
\begin{itemize}
\item on random texts on an alphabet of size 64 for $2\le m \le 10$;
\item on random texts on an alphabet of size 128 for $2\le m \le 14$;
\item on random texts on an alphabet of size 250 for $2\le m \le 18$.
\end{itemize}

\section{\label{sect-conc}Conclusion}

In this article we presented simple and though very fast improvements
 of the exact string matching algorithms of the HASH family~\cite{Lec2007}.
The new algorithms are fast for short patterns (length 2 to 18)
 on alphabet of size at least 8.
These values correspond roughly to the search of words in natural language texts.
The gain is relatively small comparing to the HASH family algorithms however
 in the era of green IT, small gains on very repetitive tasks such as looking for patterns
 can lead to large gains overall.
 
This paper constitutes a preliminary study.
It remains, at least, to conduct more experiments to see what is the expected value
 of $q$ depending on the pattern length and the alphabet size so to adjust
 the strategy for choosing between the OHASH and HASH subroutines.
The efficient use of perfect hashing for values of $q$ larger or equal to $3$
 could also improve the algorithms.

\bibliographystyle{abbrv}

\end{document}